# Collaboration in the open-source arena: The WebKit case

Jose Teixeira
University of Turku, Finland
jose.teixeira@utu.fi

Tingting Lin
University of Turku, Finland
tingting.lin@utu.fi

## ABSTRACT

In an era of software crisis, the move of firms towards distributed software development teams is being challenged by emerging collaboration issues. On this matter, the open-source phenomenon may shed some light, as successful cases on distributed collaboration in the open-source community have been recurrently reported. In this paper, we explore the collaboration networks in the WebKit open-source project, by mining WebKit's source-code version-control-system data with Social Network Analysis (SNA). Our approach allows us to observe how key events in the mobile-device industry have affected the WebKit collaboration network over time. With our findings, we show the explanation power from network visualizations capturing the collaborative dynamics of a high-networked software project over time; and highlight the power of the open-source *fork* concept as a nexus enabling both features of competition and collaboration. We also reveal the WebKit project as a valuable research site manifesting the novel notion of open-coopetition, where rival firms collaborate with competitors in the open-source community.

## Categories and Subject Descriptors

• *Software and its engineering~Open source model*   • *Software and its engineering~Programming teams*

## 1. INTRODUCTION

In an era of software crisis[1], the move of firms towards geographically-distributed, and often off-shored, software development teams is being challenged by collaboration issues. On this matter, the open-source phenomenon may shed some light, as successful cases on distributed collaboration in the open-source community have been recurrently reported [1], [2]. While practitioners move with difficulty towards globally distributed software development, there is a lack of research in academia addressing the collaboration dynamics of large-scale distributed software projects[3], [4]. In this paper, we attempt to bridge this gap by exploring the collaboration networks within the WebKit open-source project.

WebKit is an open-source project providing an engine that renders and interprets content from the World Wide Web. Its technology permeates our digital life since it can be found in the most recent computers, tablets and mobile devices sold by Apple, Google, Samsung, Nokia, RIM, HTC, and others. With more than 10 years of history, the WebKit project has brought together volunteers and firm-sponsored software developers that collaborate over the Internet by open and transparent manners while giving up the traditional intellectual property rights.

---

[1] A brief discussion on the software-crisis is provided by Fitzgerald, B. "Software Crisis 2.0." Computer 45.4 (2012): 89-91.

Previous socio-technological analysis addressing collaboration within large scale open-source software projects tend to adopt either of the two equally unsatisfactory alternatives: (1) providing thick qualitative descriptions of selected cases, thus overlooking the actors, actions and interdependent patterns of the collaborative network [5]–[7]; or (2) reducing figurational complexity to a set of quantitative indicators, thus disfiguring all practical purposes of the phenomena under investigation [8]–[10].

We opted to make our socio-technological analysis, without confining ourselves to one of the aforementioned alternatives, by analyzing how key actors and actions in the mobile-device industry affected the WebKit collaboration network over time. While addressing a previous call[11] for the advancement of methods and techniques to support the visualization of temporal aspects (e.g. pace, sequence) to represent change and evolution in ecosystems[2], we employed Social Network Analysis (SNA) over publicly-available and naturally-occurring open-source data that allowed us to re-construct and visualize the evolution of the WebKit collaboration in a sequence of networks.

The rest of this paper is organized as follows: after we briefly introduce the WebKit project, we review a series of seminal works on open-source software, and previous research addressing the open-source phenomenon by employing SNA methodological approaches. We then elaborate our methodology in details, followed by an illustration of our findings. In the end, we discuss the contributions of this paper and conclude with future-oriented remarks.

## 2. THE WebKit PROJECT

Within this section we introduce to the readers the WebKit project, giving it central significance to the research context, where we address it as a complex IT artifact that emerges and evolves as function of techno-social processes over time [12].

WebKit is an engine for browsers and other software applications. It renders and interprets content deployed on the World Wide Web where standards like HTML and JavaScript predominate. WebKit is licensed under BSD-style and LGPL licenses, thus it is freely usable for both open source and proprietary applications [13]. WebKit technologies are remarkably ubiquitous as they empower many Internet browsers (such as Apple Safari and Google Chrome) and plenty of mobile devices sold by Apple, Nokia, Samsung, RIM, HTC, Motorola, and others. Moreover, WebKit is embedded on thousands of software applications running on Windows, Mac and Linux operating-systems.

The WebKit project started as a fork of two other open-source projects: the KTML project and the KJS libraries provided by the KDE open-source community. Forking is an essential event shaping open-source communities [2], [14]; it reflects the freedom of allowing anyone to create derivative works for any purpose. In this case Apple, after deciding to enter the Internet browser market, decided to fork the KTML and KJS projects inheriting a valuable code-base for further development in accordance with their own strategy. Since its source-code (i.e. the software technology blueprint) was published by Apple, it has been further developed by non-affiliated open-source developers (i.e. from the KDE community) and others from firms like Apple, Google, Nokia, RIM, Igalia, Intel and Samsung. Since Apple's WebKit debut, the overall project was once again forked in 2010, leading to the creation of the WebKit2 project for a more platform-independent version. More recently, Google announced that it had forked core components of WebKit to be used in future versions of its browsers[3]. Figure 1 illustrates the forking within the WebKit history.

---

[2] Basole, R. employs the ecosystems term as a complex network of companies interacting with each other, directly and indirectly, to provide a broad array of products and services. Thus the ecosystem metaphor can also be applied in the WebKit project.

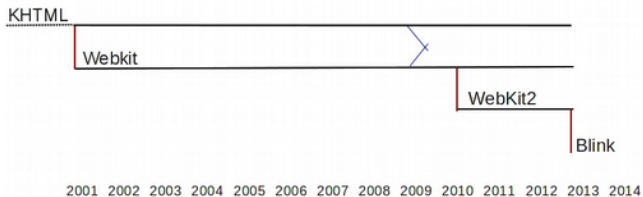

**Figure 1: Forks within the WebKit project**

Resembling the peer-reviewed mechanisms employed in the academia, the WebKit coding policy distinguishes and empowers different actors, including regular contributors, committers and reviewers. Similar to other open-source communities, the WebKit project is also based on a high level of meritocracy, where software developers are ranked by their prior contributions to the community [2], [15] that are evaluated by their peers within the network.

Even if WebKit has attained a remarkable ubiquity, the project has been rarely addressed by the academia. In this research, we strive to study the collaboration dynamics in the open-source community; WebKit is then an ideal case field given its highly collaborative and networked characteristics.

## 3. RELEVANT LITERATURE

The open-source phenomenon has attracted steady attention from multi-disciplinary scholars in the last decades [2], [5], [14], [16]. To illustrate the growing academic relevance of the open-source phenomenon, we observed that many prominent academic outlets, including "Research Policy" ,"IEEE Network", "IEEE Software", "Management Science" and the "Journal of the Association for Information Systems", have recently published special issues on open-source software. Several recent and comprehensive literature reviews have addressed the open-source phenomenon [17]–[20]. And the phenomenon keeps evolving from the earliest purist views focusing on freedom [21], to a newer perspective considering open-source as an alternative and viable business [14], [22].

Few scholars have leveraged the network perspective and the SNA approach to study the open-source phenomenon. However, there are some notable exceptions [8], [10], [23] who have based their network analysis on metadata from public source-code repositories and/or email data in bug-fixing contexts. We also conducted SNA in this research; however, unlike most of the above-mentioned research with cross-sectional analysis of static networks, we adopted a longitudinal view as we are more interested on how the collaboration network evolves over time.

---
[3] Google announcement of Blink, a WebKit project fork is available at http://blog.chromium.org/2013/04/

Moreover, this research departs from the prior research with a new aim to understand how mobile device vendors collaborate on the open-source arena. Rather than analyzing solely the social network of the WebKit community, we also acknowledged key actors and actions on the higher level of mobile-device industry, seeking to understand how key exogenous events in the industry have affected WebKit and its social network. To sum up, rather than extracting quantitative indicators from the collaborative network by solely looking at IT artifacts[4], we also look at its surrounding industrial environment seeking for understanding on how different happenings on the industry shaped the collaboration network developing the same IT artifacts.

## 4. METHODOLOGY

In this section we will elaborate on our research design and methodological details. Without ever leaving our labs, and by looking at naturally-occurring data publicly available on the Internet, our methodology combines the screening of key happening in the mobile devices industry with a computer-based method of SNA.

We started by screening, by ethnographic manners, publicly available data such as company announcements, financial reports and specialized-press that allowed us to review immense online information pertaining to the competitive mobile-devices industry; therefore, we were able to study the insight of the industrial context. After attaining a better understanding of the the competitive dynamics of the mobile-devices industry, we later started extracting and analyzing the social network of the WebKit community leveraging SNA [24], [25], which is an emergent method widely established across disciplines of social sciences in general[25]–[28] and information systems in particular [10], [29], [30] .

We first built the social network matrices with UCINET[31] based on the WebKit project change-log. In the analysis, we focused on the visualization of the collaboration network, which evolves over time, to reveal dynamics among the WebKit software developers. We then attempted to understand the visualized networks with our previously acquired tacit understanding from the competitive mobile-devices industry.  The visualization, together with a deeper understanding of the phenomenon under investigation, corresponds to the notion of figuration [32]as pointed out by some prior multi-disciplinary studies [33]–[36]. We provide more details of our data collection and analysis in the following sections.

### 4.1    Data-collection

Our screening of public and natural-occurring data available on the Internet followed the general ethnographic principles that have been extensively established in social sciences and information systems [37], [38]. Specifically, we have reviewed relevant firm's public announcements, publicly available financial reports, news from specialized press and discussions in forums and blogs. Our empirical materials span the time period from September 2006 until April 2013, and all are freely available to the public on the Internet.

After acquiring a deeper understanding from the competitive dynamics of  mobile-devices industry, we also conducted SNA which allows us to depict overall pictures of the collaborative dynamics among different developers in the WebKit project. The input data of SNA is based on different source-code versions of the WebKit project. Our last compilation was filed on 3rd April 2013, which comprises ca. 1.4GB. From the version-control change-log documentation[5], we extracted basic information as input for the SNA, including each developer's email address and the time stamp when he/she made a change to a specific file (see Figure 2). We then connect the developers who work on the same file, and construct a network of collaboration activities among all the developers. With the visualization of the collaboration network over time, we aim to understand the evolution of the code-based collaboration with a lens of social structure. We will describe the details of our data analysis in the following section.

---
[4] In our case WebKit source-code and it's version-control-system

## 4.2 Data-analysis

While screening the competitive dynamics of mobile-devices industry we selected key events from the industry regarding open-source software in general and the WebKit project in particular. We started with a chronological approach; however, we went back and forth in the dynamic history of the mobile-devices industry, trying to make sense of our online observations. Our practice-accumulated skills, regarding software development, open-source software and software version control systems, dealing with very specific concepts and terminologies, revealed to be essential for sense-making of the collected ethnographic material. We have identified a set of endogenous and exogenous events that, according to our interpretations, could have impacted the evolutionary dynamics of the WebKit project (see Table 1). These major events give us a more clear history line to understand the evolution of this project, as well as the industrial context in which it is embedded.

**Table 1. Key selected events within WebKit**

| Date | Event |
| --- | --- |
| Jun 2001 | WebKit started within Apple as a fork of KHTML and KJS open-source projects. |
| Sep 2006 | Apple, forced by the open-source community, published WebKit source-code in a public repository. |
| Jun 2007 | Apple released 1st generation of iPhone |
| Sep 2008 | Google launched Chrome and Android |
| Jun 2009 | Nokia and Intel Announced Strategic Relationship |
| Feb 2011 | Nokia and Microsoft formed a broad strategic partnership. Intel searched for new partners for Meego. |
| Jul 2012 | The patent war broke out between Apple and Samsung, and their hostilities reached climax with the first trial in U.S. $1.049 billion in damages. |
| Apr 2013 | Google announced to fork WebKit's core components, just 1 month after Apple registered WebKit as its trademark. |

The qualitative ethnographic efforts, conducted prior to and during our computerized SNA, revealed to be fundamental while analyzing the WebKit social network evolutionary dynamics. To prepare for the SNA, the identified industry-events were used as partitions on the whole period of the project history since 2006[6]. We then applied SNA and constructed the collaboration network of developers in each partitioned time slice. In this way, we are able to assess how the collaboration network has evolved over time in response to the exogenous events in the industry. Specifically, the input of SNA was based on developers' active contributions to the WebKit source-code from 1 September 2006 till 3 April 2013. These contributions were documented in the publicly available WebKit change log produced both by the WebKit committers (i.e. the ones with read-write access to the project repository) and the WebKit reviewers (i.e. the ones with a final word on what stays in or out of the project blueprints). Figure 2 shows a sample of the change log to illustrate how the collaboration network is identified and constructed.

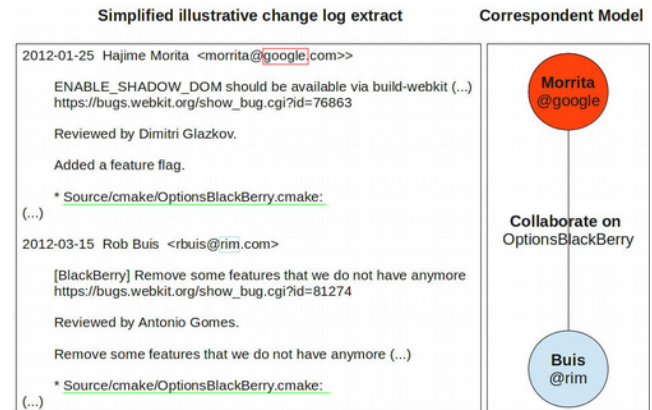

**Figure 2: Modeling the WebKit change log**

The log was parsed, validated and processed with the Python programming language, tracing back all collaborations in a period of almost 7 years since September 2006. By 3 April 2013, when Google forked the WebKit project to create Blink[39], we could identify 445 nodes and 2169 edges, forming a complex mesh in which 445 software developers have worked together.

The collaborative network during a certain time slice can be formally defined as:

$$G_t = (V, A_v, E)$$

Where:

$V$ = A set of nodes representing the developers contributing to the WebKit open-source software project

$E$ = A set of edges, identifying the connections between two developers if they have worked on the same software source-code file.

$A_v$ = A set of nodes-attributes, capturing each developer's company affiliation. This information is extracted from the email address of each developer.

Based on this definition, we used UCINET[40] to build the network matrices. Various numeric network measures have been established in SNA: For example, eigenvector-centrality [41], [42] degree-centrality and betweenness-centrality [25] reveal the importance of a node in a network. Other aspects of a network can also be manifested with important measures such as network-density [24], cluster coefficients [43], strength of ties [44], etc. However, as our SNA goes hand-in-hand with a interpretivist ethnography on the competitive mobile industry, the visualization of network graphs is sufficient to naturally uncover the history line and the dynamics of collaboration in a qualitative and straight-forward way. The visualization of social networks has been widely used by scholars [33]–[36], but few studies have explored the time dimension to observe how networks evolve [45]. We used the software Visone [46] to visualize a sequence of networks according to the established time slices partitioned by the major events, and interpreted the network evolution with understandings generated from the previous collection of rich qualitative material capturing the competitive dynamics of the mobile-devices industry.

---

[5] A book on the practice of version control systems in the open-source community is freely available on the Internet at http://svnbook.red-bean.com/. It might contribute to a better understanding on how our data was collected.

[6] Although the WebKit project started in 2001, in the raw-data we can access, the earliest change on WebKit source-code is only documented in 2006.

For a better understanding on the industry level, we opt to focus on the network of developers from major mobile device vendors involved in the WebKit project. The selection of these major vendors is based on a prior public-report by Bitergia on WebKit collaboration [47], where the 10 most active organizations have been identified on the development of the WebKit project, including Apple, Google, Nokia, Rim, Igalia, Intel, Samsung, Univ. Szeged (Inf), Adobe, and Torchmobile. Therefore, we highlighted these 10 vendors with different colors in the visualized networks, and marked other developers' affiliation as "other" and in gray color. It is worth noticing that most software developers within WebKit are non-affiliated developers without explicit firm-sponsorship; therefore, most of "other" developers are independent contributors.

## 5. FINDINGS

In this section, we illustrate our findings with network visualizations showing the evolution of the collaboration network throughout the development progress of WebKit software source-code.

Our visualizations (Figure 3-6) facilitate an intuitive understanding on how key players in the mobile-devices industry collaborate in the open-source arena. The first visualizations (Figure 3-4) capture the early development of the WebKit project; while our last visualizations (Figure 5-6) capture the hyper-collaborative nature of the WebKit project during the last four years, when it started empowering our computers and mobile-devices in a larger scale. Using Visone [46], we also visualized the centrality of each developer by differentiation on node size, i.e. the larger the node is, the more central the represented developer acts in the community. The value of centrality depends on the number of adjacent nodes that a node is connected with. Therefore, the higher a developer's centrality is, the more active he/she is in collaborating with others.

Our first network-visualization, i.e. Figure 3, depicts the collaboration on the WebKit project from 1 September 2006 (i.e. when apple first published WebKit source-code) to 29 June 2007 (i.e. when Apple released the first generation iPhone leading to the emergence of millions of mobile-devices powered by WebKit). From this visualization of early WebKit history, we can observe four developers from Apple who collaborate only among themselves, segregated from others in the WebKit community; while one Apple-affiliated developer acts as a bridge to the rest of WebKit community. Interestingly, the latter Apple developer doesn't have any connection with other four colleagues during this particular period in the project. Although the total number of nodes is relatively small at this early stage, there is no isolates despite the evident segregation between the two sub-networks.

Our second network visualization, Figure 4, captures a thriving phase of collaboration within the WebKit project from 29 June 2007 (i.e. the release of the first iPhone) to the end of September 2008 (i.e. the month that Google launched Chrome and Android platforms integrating WebKit). Although the number of Apple affiliated developers remains the same, the project has attracted increasing participation among non-affiliated developers. Meanwhile, one developer from Torchmobile emerged in the network. In addition, the density from the network has increased compared to the last visualization in Figure 3.

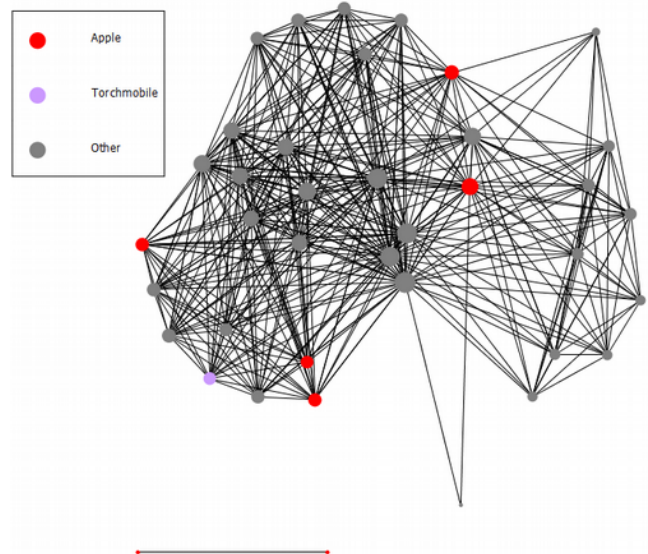

**Figure 4: WebKit and KHTML join forces**

This eye-opening expansion of the WebKit community and intensified collaboration can be partially explained by the unforking of KDE's KHTML and [48]. It indicates that after years of split development of WebKit and KHTML (though with code exchanges to integrate on both sides), Apple and KHTML developers have decided to increase collaboration and many KHTML developers have become reviewers and submitters for the WebKit source-code repository, and vice versa.

Our third network-visualization, Figure 5, demonstrates the later phase of the WebKit project, starting from the end of September 2008 (i.e. the launch of Chrome and Android) to 3 February 2011 (i.e. when Nokia and Microsoft announced a strategic partnership leaving alone Intel with the Meego platform [49], [50].

During this phase, considering the companies' participation on the WebKit development, Apple has lost its unique central-role, and shares network-centrality with Google, Samsung and Igalia. On the other hand, RIM and Nokia, adopting WebKit within their latest mobile platforms, remains in periphery with observable separation from the most central players.

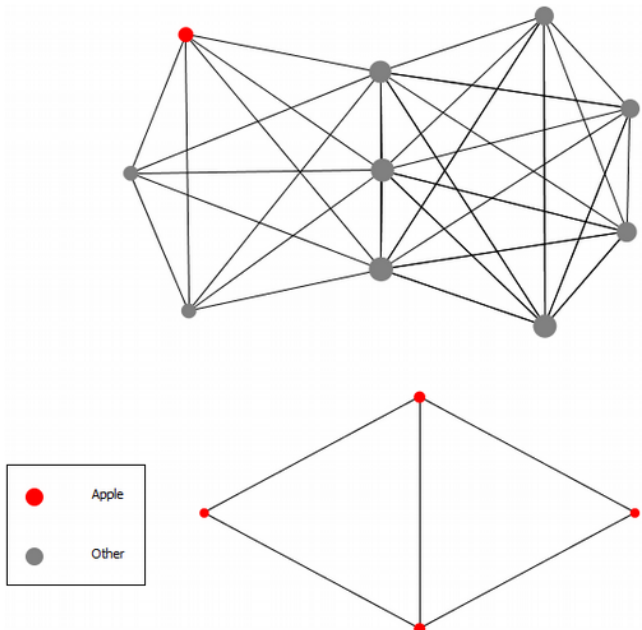

**Figure 3: Visualizing the WebKit bootstrap**

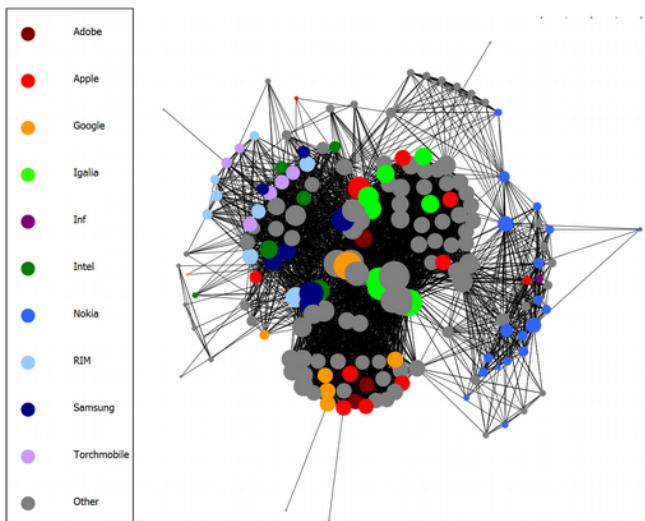

**Figure 5: Mass collaboration**

However, thanks to our previous qualitative ethnographic work we must highlight that this visualization must be interpreted carefully, since Igalia, a Spanish firm specialized on open-source software development services, has been working on the projects during this period. Providing software services to many major firms, Igalia often represents the interests of Nokia and Intel on the aemo and Meego platforms [51]. Therefore, given Igalia's central position in the network, we cannot conclude the peripheral role of Nokia and Intel despite their network position. Nevertheless, the clear separation between Nokia and Intel, who are former partners in the Meego project [52], [53], in the network is consistent with the breakage of cooperation between the two companies, due to the new partnership strategy Nokia adopted at that time.

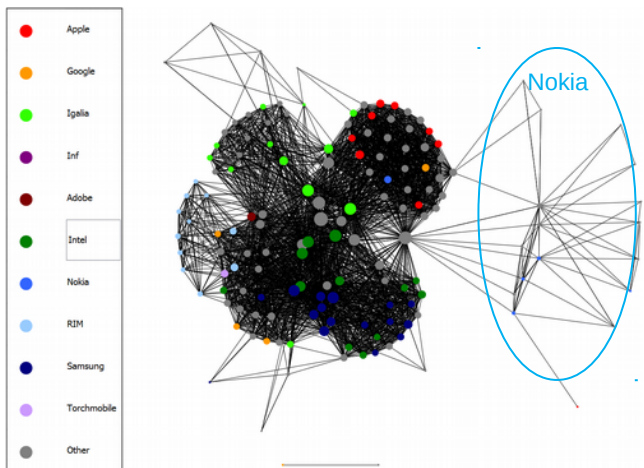

**Figure 6: Patent-wars, trademarks and forking**

Our last network-visualization, Figure 6, illustrates the latest phase of the WebKit project from the end of 3 February 2011 (i.e. Nokia and Microsoft's announcement of a strategic partnership) to 3 April 2013 (i.e. Google forks the WebKit core creating the Blink project).

Similar to Figure 5, in Figure 6 we can also observe that contributors sponsored by Nokia and Intel are on opposite sides of the network, reflecting the lack of collaboration between those two firms in the WebKit project. This lack of collaboration increased as Nokia became increasingly dependent on Microsoft software to power their devices. Therefore, Nokia become even more peripheral in the open-source community, which is also visible in the visualized network. As compared to Figure 5, here the blue nodes representing Nokia developers have significantly decreased in size and moved further away from the central groups.

Comparing Apple and Samsung's roles in the networks shown in Figure 5 and Figure 6, we also attain interesting findings: Even if Samsung and Apple are involved in expensive patent wars in the courts [54] and stopped collaborating on hardware components [55], their contributions remained strong and central within the WebKit open-source project. However, the distance between the two groups has indeed increased as the rivalry has upgraded since the patent wars in 2012.

Additionally, across all visualizations, non-affiliated developers, who are often volunteers without firm-sponsorship, are more central within the WebKit collaboration network than developers affiliate with the 10 organizations we highlighted according to the previously mentioned Bitergia study [47].

## 6. DISCUSSION

Before discussing the contributions and implications of our study, it is important to mention that this research is entirely based on naturally-occurring data available to the public on the Internet. Thanks to WebKit's strict policy for committers and reviewers, our data set was extremely clean, facilitating a smooth data extraction ahead of SNA. Thus, our data cleansing efforts were minimal, contrasting with prior research reporting enormous difficulties in the collecting, cleaning and screening of open-source projects data[8], [10], [23].

### 6.1 Academic contributions

Our findings seem to integrate with a variety of theories on management, cooperation and innovation in networked communities. Perhaps one of the most interesting ones, that explains features from the evolutionary collaborative dynamics of the WebKit project is a management theory on the paradox of firm investment in open-source software [56] stating that in a scenario of pooled R&D development, the firms adopting open source components have four common characteristics:

- there is pre-existing open source code being developed without the intervention of the focal firms;
- the "buy vs. build" decision to use external innovation is made easier because the code was "free";
- the firms were willing to contribute back to the existing projects on an ongoing basis, to assure that the technology continued to meet their respective needs, to maintain absorptive capacity, and to avoid discouraging external innovators;
- the firms could continue to yield returns for internal innovation by combining the internal and external technologies to make a product offering that was not directly available through open source.

In this WebKit case, we can observe that the most active firms contributing to the project exhibited all of the previous mentioned characteristics, while the more peripheral firms failed to meet the third characteristic. This last group of firms were clearly more interested in integrating WebKit into their technological pools without strategically contributing back[7].

---

[7] Coincidence or not, firms that played a more central role in the WebKit project such as Google, Apple and Samsung were by 2013 the leaders of the mobile-devices industry. While more more peripheral firms such as RIM and Nokia lost market-share.

Another theoretical contribution that emerged from our approach highlights the power of the open-source *fork* concept as a nexus enabling both features of competition and collaboration. As previously mentioned, *fork* reflects the open-source freedom of allowing anyone to create derivative works. A *fork* divide a community in two, the simple existence of a threat of a *fork* have significant implications within a previously united community. As a form of schism, any developer have the freedom of leaving the community, with a copy of the existing code-base and further develop the project by its own manners. It was argued before that [57] that *fork* serves as an invisible hand of sustainability ensuring that the code-base remains open and best fulfills the needs of the community it lives on. The occurrence of several forks on the initial WebKit code-base (see Figure 1) is better understood with prior work [58] that identifies the need of porting a program to a new hardware or software architecture as a driver of forking[8].

In the WebKit case, *fork* enabled a set of networked collaboration features: The existence of an existing code-base reduced the barriers to entry of firms seeking to integrate Internet-browsing technologies into their digital platforms. The initial WebKit code-base was then forked several times as more and more firms were interested in porting the "program" into heterogeneous hardware/software stacks. On other hand, the threat of a *fork* stimulated a collaborative sense of community [59] and the setup of basic norms and values [60] unifying the community against possible break-up forces. All this in a scenario of pooled R&D where costs and governance are shared within a collaborative community[56].

Even if the initial goal of this research was to study collaboration in the WebKit project, we identified that *fork* also enables a set of competition features: First of all, even if *fork* facilitates the commoditization of technology that can be copied and ported to architecturally different products, in the WebKit case this only concentrated a small effort of the "whole product" offering from many of the involved firms. Firms relying on WebKit source of innovation, kept differentiating both while porting it to their own architectures and in other areas of their computer-based platform/ecosystem. Moreover, firms exhibit competition when recruiting talented open-source developers or when sourcing from open-source service providers[9]. Besides competing for talented labor needed for developing such a large-scale open-source *fork*, firms also compete for abortive capacity[56], [62], technological learning [6] and organizational learning [6], [63]. With the previous mentioned reduced barriers to entry there is an increased risk of free riding[64], innovators must master the open-source community project for better guiding its development according their own interests while being aware that copycats[10] can always *fork* their contributions.

Our research witnessed a peculiar extent of collaboration between rival firms from the evolving network, moreover we recognized *fork* as a nexus enabling both features of competition and collaboration, leading us with the proposition that the open-source community can also be a great arena to observe the phenomenon of coopetition[65], [66]. However, we were not able find published Management or Information Systems literature exploring coopetition features in the open-source arena[11], an area that we will further explore while proposing already a neologism:

Open-coopetition: A portmanteau of cooperative competition in the open-source arena, where R&D is jointly performed by competing firms by open-source manners, giving-up authorship-granted intellectual property rights for maximizing both blueprints transparency and collaborative benefits.

## 6.2    Implications for practice

We shed lights on the potential of visualizing the evolutionary collaborative dynamics in R&D projects, especially for practitioners dealing with large-scale and networked productions. Different stakeholders in large-scale open-source software projects could gain strategic and operational benefits: For software developers, our methods empower them with better understanding on the overall network to improve development processes. For users, adopters and integrators, we can depict the project evolution for thorough assessments of its sustainability and dynamics when reacting to exogenous events. And for investors, clarifying the network dynamics can improve the forecast of product attractiveness and future growth.

We also provided a rich description on how hight-tech giants collaborated with rival-competitors in the WebKit project by open-source manners. Given the current financial success of the high-tech firms with a more central role in WebKit development (i.e. Apple, Google, Samsung), R&D managers are reminded once again for the dangers of ignoring open-source software as a external source of innovation.

## 7. CONCLUDING REMARKS AND FUTURE RESEARCH

In this paper, we attempt to provide a better understanding of how key players of mobile-device industry collaborate in the open-source arena, by investigating the development of the WebKit project. We combined an ethnographic approach and network visualization supported by SNA. Our findings show the explanation power of such mixed-methods on the meanings of network dynamics and highlight the power of the open-source *fork* concept as a nexus enabling both features of competition and collaboration.

For future research, we aim to further theorize our findings integrating the notion of open-coopetition, in a quest for better understanding how firms collaborate with competitors in the open-source arena. We will further explore the concept of forking as we will align our research journey with an ongoing development of the WebKit project, assessing how the current WebKit social network will be affected by Google's recent decision to fork the WebKit project.

## 8. AKNOWLEDGEMENTS

The idea of this research project surged by pure serendipity at the *Inforte seminar on Big Data and Social Media Analytics* by Sudha Ram and Matti Rossi. The researchers thank the financial support from the *Fundação para a Ciência e a Tecnologia* (grant SFRHBD615612009) and *Liikesivistysrahasto* (grant 3-1815). Acknowledgements also for *Lero - the Irish software engineering research centre* were part of this research was conducted. Special thanks to Jari Salo, Reima Suomi, Sarah Beecham and Gregorio Robles for early comments on manuscripts. A last word to the

---

[8] I.e Google argued that the complex architectures of WebKit were slowing down the collective pace of innovation when announcing its Blink fork of WebKit.

[9] According Agerfalk and Fitzgerald open-source service providers are typically SMEs [61]

[10] Even if copycats term is often used in management to refer to free-riders in the emerging economies, in the open-source world it refers to firms that integrate open-source technologies without contributing back up-stream to its development.

[11] An interesting pharmaceutical article bridging open-source and coopetition by Munos, B. "Can open-source R&D reinvigorate drug research?" Nature Reviews Drug Discovery 5 (2006): , 723-729

WebKit developers for developing cool, open and research-friendly technologies.

More methodological details, data, high-resolution visualizations and source-code at http://users.utu.fi/joante/WebKitSNA/.